# Surprisingly large anomalous Hall effect and giant negative magnetoresistance in half-topological semimetals

Yanglin Zhu[1+], Cheng-Yi Huang[2+], Yu Wang[1], David Graf[3], Hsin Lin[4], Seng Huat Lee[1], John Singleton[5], Lujin Min[1], Johanna C. Palmstrom[5], Arun Bansil[2], Bahadur Singh[6*], and Zhiqiang Mao[1*]

[1] Department of Physics, Pennsylvania State University, University Park, PA, 16802

[2] Department of Physics, Northeastern University, Boston, USA, 02115

[3] National High Magnetic Field Laboratory, Tallahassee, FL, 32310

[4] Institute of Physics, Academia Sinica, Taipei 11529, Taiwan

[5] National High Magnetic Field Laboratory, Pulse Field Facility, Los Alamos National Laboratory, Los Alamos, NM, 87545

[6] Department of Condensed Matter Physics and Materials Science, Tata Institute of Fundamental Research, Mumbai 400005, India

**Abstract**

Large intrinsic anomalous Hall effect (AHE) due to the Berry curvature in magnetic topological semimetals is attracting enormous interest due to its fundamental importance and technological relevance. Mechanisms resulting in large intrinsic AHE include diverging Berry curvature in Weyl semimetals, anticrossing nodal rings or points of non-trivial bands, and noncollinear spin structures. Here we show that a half-topological semimetal (HTS) state near a topological critical point can provide a new mechanism for driving an exceptionally large AHE. We reveal this through a systematic experimental and theoretical study of the antiferromagnetic (AFM) half-Heusler compound TbPdBi. We not only observed an unusual AHE with a surprisingly large anomalous Hall angle $\Theta^H$ (tan$\Theta^H \approx 2$, the largest among the antiferromagnets) in its field-driven ferromagnetic (FM) phase, but also found a distinct Hall resistivity peak in the canted AFM phase within a low field range, where its isothermal magnetization is nearly linearly dependent on the field. Moreover, we observed a nearly isotropic, giant negative magnetoresistance with a magnitude of ~98%. Our in-depth theoretical modelling demonstrates that these exotic transport properties originate from the HTS state. A minimal Berry curvature cancellation between the trivial spin-up and nontrivial spin-down bands results not only in an extremely large AHE, but it also enhances the spin polarization of the spin-down bands substantially and thus leads to a giant negative magnetoresistance. Our study advances the understanding of the interplay between band topology and magnetism and offers new clues for materials design for spintronics and other applications.

+These two authors equally contribute to this work

*emails: bahadur.singh@tifr.res.in, zim1@psu.edu

Large intrinsic anomalous Hall effect (AHE) in topological semimetals has been a subject under intensive studies. Unlike the conventional Hall effect, which is caused by the Lorentz force under an external magnetic field, the intrinsic AHE stems from the net Berry curvature $\Omega(k)$ of the band structure [1-3]. $\Omega(k)$, which describes the geometry of the Bloch wavefunctions, is determined by the topology of the band structure. When a longitudinal electric field is applied, $\Omega(k)$ imparts transverse velocity [$\propto \boldsymbol{E} \times \boldsymbol{\Omega}(k)$] to Bloch electrons and thus results in the AHE. To observe intrinsic AHE, time-reversal symmetry (TRS) is required to be broken. Therefore, large intrinsic AHE is usually expected in magnetic materials with Berry curvature hot spots near the Fermi level. Prior studies have found that several types of magnetic topological materials can exhibit large intrinsic AHE. These include ferromagnetic (FM) Weyl semimetals such as $Co_3Sn_2S_2$ [4,5] and $Co_2MnGa$ [6,7], which host Weyl nodal crossings that lie close to the Fermi level and carry a diverging Berry curvature. With the TRS broken by ferromagnetism, the Berry curvature contribution from the Weyl nodes with opposite chirality will not cancel out, resulting in a large AHE. Strength of the AHE is usually characterized by the intrinsic anomalous Hall angle (AHA) $\Theta^H$, where $\Theta^H = \tan^{-1}(\sigma_{yx}^{AH}/\sigma_{xx})$, and $\sigma_{yx}^{AH}$ and $\sigma_{xx}$ are the anomalous Hall conductivity and longitudinal conductivity, respectively. $Co_3Sn_2S_2$ and $Co_2MnGa$ both exhibit large AHA with $\tan\Theta^H = 0.2$ ($Co_3Sn_2S_2$) [4] and 0.12 ($Co_2MnGa$) [8]. Antiferromagnetic (AFM) half-Heusler materials such as GdPtBi [9] and TbPtBi [10] also display a large intrinsic AHE with $\tan\Theta^H$ values as large as 0.16-0.76 [9-12]. Although these materials harbor Weyl nodes induced by magnetic fields, their large AHE does not originate from Weyl nodes but arises from the large net Berry curvature produced by the anticrossing of spin-split bands near the Fermi level [9,12]. Besides the large AHE caused by the presence of anticrossing Weyl nodes or bands, recent studies show that gapped nodal rings can also generate extremely large AHE [13,14]. This was experimentally demonstrated in FM Heusler $Co_2MnAl$ [14], whose AHA reaches a record value with $\tan\Theta^H = 0.21$ at room temperature. In addition, the non-collinear AFM structure can lead to a large Berry curvature, resulting in a large intrinsic AHE [15,16]; this has been seen in a range of AFM topological materials with broken TRS, such as $Mn_3Sn$ [17] and $Mn_3Ge$ [18].

In this paper, we report a surprisingly large AHE in the Pd-based half-Heusler compound TbPdBi. Its $\tan\Theta^H$ value reaches ~ 2, which is the largest among all known magnetic topological materials. Our detailed theoretical analysis suggests that such an extremely large AHE originates from a half-topological semimetal (HTS) state. HTS is a long-sought topological state in materials and can be viewed as a topological version of the half-metallic state in which electrons conduct in only one spin channel, while the other spin channel is insulating [19-21]. Since such a state could generate low power consuming spin current, it holds great promise for applications in topological quantum spintronic devices [22]. Moreover, it has been theoretically shown that a gap opening at the non-trivial band crossing points transforms the HTS into a quantum anomalous Hall insulator [23,24]. These exotic properties have inspired extensive interest and a variety of material systems have been predicted to be HTS, such as two-dimensional(2D) MnN [25], PrOBr [26], and $PtCl_3$ [24]; quasi-1D $X_2RhF_6$ (X=K, Rb, Cs) [27] and $XYZ_3$ (X=Cs, Rb, Y=Cr, Cu, Z=Cl, I) [28]; 3D $MF_3$ (M=Pd, Mn) [29], $LiV_2F_6$ [30], etc. The Dirac/Weyl points or nodal rings in all these materials are comprised of spin-polarized bands [21,30,31]. However, these theoretical predictions are still awaiting experimental verifications. Our work here demonstrates that TbPdBi hosts a unique HTS in proximity to a topological critical point. Such a peculiar HTS in TbPdBi results in not only an unusually large AHE but also leads to a giant negative magnetoresistance. Additionally, we find that the Hall resistivity of TbPdBi exhibits a distinct anomalous peak in the low field range, where its isothermal magnetization nearly linearly depends on field and this behavior can be understood in terms of the Berry curvature enhancement induced by spin canting.

Single crystals of TbPdBi were grown using the Bi-flux method (see Methods). They exhibit semi-metallic behavior, manifested by the broad peak in the temperature dependence of resistivity $\rho_{xx}$ around 50K (see supplemental Fig. S1a). The magnetic susceptibility ($\chi$) measurements (Fig. S1a) show its AFM state with $T_N$ = 5.2 K. $\rho_{xx}$ exhibits a steeper drop below $T_N$, indicating electronic transport coupled with magnetism. We performed a Curie-Weiss (CW) fit for its temperature dependence of susceptibility $\chi(T)$; the best fit was obtained in the temperature range of 100-300K (Fig. S1b) which yields the effective magnetic moment ($\mu_{eff}$) of 9.4 $\mu_B$/Tb, consistent with a prior report [32].

From Hall resistivity $\rho_{xy}$ measurements under high magnetic fields, we observed exceptionally strong AHE in TbPdBi. Figure 1a shows the $\rho_{xy}$ data measured up to 31T at various temperatures. The most significant feature of the $\rho_{xy}$ data of TbPdBi is that it exhibits a striking peak. This peak not only occurs below $T_N$ (= 5.2K) but also extends to temperatures above $T_N$. The peak field is ~ 5T below $T_N$, and slightly shifts to high field with increasing temperature. $\rho_{xy}(B)$ gradually evolves into a linear field dependence after exhibiting a peak and this occurs above 15T for $T < T_N$, at higher temperatures (e.g. 15K & 20K), due to the shift of the peak to higher field, the linear trend develops at higher fields, with the linear slope remaining similar to those of low temperatures. Such a distinct peak in $\rho_{xy}$ has never been observed in any other half-Heusler compounds, full-Heusler antiferromagnets, or conventional ferromagnets. As noted above, prior work has shown isostructural half Heusler compounds (Gd/Tb)PtBi also exhibits large AHE [9-12]; their $\rho_{xy}$ data also display anomalous peaks near 4.5T where their AHAs achieve maxima with $\tan\Theta^H$ = 0.16-0.76 [9-12]. However, the $\rho_{xy}$ anomalous peak of (Tb/Dy)PtBi is far weaker than that of TbPdBi. For comparison, we have added the $\rho_{xy}$ data of TbPtBi at 1.7K to Fig. 1a. While its weak $\rho_{xy}$ peak near 4.5T can be clearly resolved when the data is zoomed into the field range of 0-9T (see Fig. 1(c) in ref. [12]), it is hardly discernible when this data is plotted together with the data of TbPdBi in the field range up to 31T as shown in Fig. 1a, suggesting that the replacement of Pt by Pd leads to essential changes to the band structure, as discussed below. In the field regime where $\rho_{xy}$ exhibits an anomalous peak, the longitudinal resistivity $\rho_{xx}$ displays a drastic decrease, followed by a saturation trend in the high field range where $\rho_{xy}$ evolves into a linear field dependence, as shown in Fig. 2b which presents representative $\rho_{xx}$ and $\rho_{xy}$ data at 4K. Note that the anomalous $\rho_{xy}$ peak near 5T is not caused by the $\rho_{xx}$ component which might not be removed from the $\rho_{xy}$ data's anti-symmetrizing process; this can be seen clearly from the raw data of $\rho_{xy}$ and $\rho_{xx}$ shown in supplementary Fig. S2 which shows the $\rho_{xx}$ component in $\rho_{xy}$ is negligibly small. To find whether the unusual field dependences of $\rho_{xy}$ and $\rho_{xy}$ arise from a magnetic transition, we measured the magnetization of TbPdBi up to 35T at various temperatures (see Methods) and added the data from these measurements to Fig. 1b. From these data, we find a spin-flop transition near 15T and the saturated magnetic moment of $Tb^{3+}$ in the FM phase is ~8.5$\mu_B$/f.u. at 0.57K, comparable with the effective magnetic moment extracted from the CW fit ($\mu_{eff}$ ~ 9.4 $\mu_B$/Tb). The unusual field dependences of $\rho_{xx}$ and $\rho_{xy}$ are indeed coupled with such a magnetic spin-flop transition. The $\rho_{xy}$ peak as well as the sharp drop of $\rho_{xx}$ are present in the canted AFM state, while the linear field dependence of $\rho_{xy}$ and the $\rho_{xx}$ saturation behavior occur in the polarized FM phase. These observations imply that the spin-flop transition leads to an electronic structure transition.

As to be shown below, our theoretical calculations shows that such a spin-flop transition gives rise to a HTS state.

Given that the remarkable $\rho_{xy}$ peak is present in the CAFM state, its origin is most likely associated with Berry curvature induced by noncolinear spin structure. In general, $\rho_{xy}$ of magnetic systems can be described by $\rho_{xy} = R_0 B + R_S M + \rho_{xy}^T$, where the first term is normal Hall contribution ($R_0$, the normal Hall coefficient), the second term represents the anomalous Hall resistivity linearly coupled with magnetization $M$ ($R_S$, the anomalous Hall coefficient), and the last term $\rho_{xy}^T$ is the anomalous Hall resistivity arising from Berry curvature effects. The remarkable $\rho_{xy}$ peak of TbPdBi present in the canted AFM state suggests it involves a significant component of anomalous Hall resistivity $\Delta\rho_{xy}^{AH}(=\rho_{xy} - R_0 B)$. To find how the magnetization contributes to $\Delta\rho_{xy}^{AH}$, we have plotted $\Delta\rho_{xy}^{AH}$ as a function of magnetization in supplementary Fig. S3 (Note that $\Delta\rho_{xy}^{AH}$ is obtained by subtracting the normal Hall contribution, i.e. the dashed line in Fig. 1a which is inferred from the high-field linear field dependence of $\rho_{xy}(B)$). We find $\Delta\rho_{xy}^{AH}$ strongly deviates from linear dependence on magnetization and exhibits a striking peak, indicating the anomalous Hall resistivity of TbPdBi involves a significant contribution of $\rho_{xy}^T$. The maximal $\Delta\rho_{xy}^{AH}$ near 5T is ~ 0.67 mΩ cm (supplementary Fig. S4a), almost 5-15 times larger than that in (Gd/Tb)PtBi [9,12]. From $\Delta\rho_{xy}^{AH}$, we derive anomalous Hall conductivity $\sigma_{yx}^{AH}$ using $\sigma_{yx}^{AH} = \Delta\rho_{xy}^{AH}/(\rho_{xy}^2 + \rho_{xx}^2)$ and AHA as shown in supplementary Fig. S4b and Fig. 1c. The maximal value of tan$\Theta^H$ is ~ 2 at about 20T and 10K (Fig.1c), which is surprisingly large as compared to other magnetic topological materials, as shown in Fig. 1d, which plots tan$\Theta^H$ versus $\sigma_{yx}^{AH}$ for TbPdBi and other magnetic topological materials showing large AHE. Such an extremely large AHE seen in TbPdBi is unlikely induced by extrinsic mechanisms such as skew scattering or side jump, since their induced Hall angle is usually a few percent [33,34]. The intrinsic mechanisms due to Berry curvature should play a key role in generating such an unusually large AHE in TbPdBi as discussed below. Previous studies have shown Skyrmion magnetic lattices could give rise to an anomalous peak in $\rho_{xy}$ (i.e. topological Hall effect [35]). Compared to the topological Hall effect of a prototype Skyrmion system MnSi where anomalous Hall resistivity jump (i.e. $\rho_{xy}^T$) induced by the spin-texture is less than 0.04 μΩ cm [36], our observed maximal $\Delta\rho_{xy}^{AH}$ value of ~0.67 mΩ cm in TbPdBi is four orders of magnitude larger. This implies that the extremely large AHE in TbPdBi should have a unique mechanism. As to be shown below, it is indeed associated with a peculiar HTS state.

The steep decrease of $\rho_{xx}$ in field regions where $\rho_{xy}$ exhibits an anomalous peak (Fig. 1b) suggests that TbPdBi exhibits large negative magnetoresistance. Although we observed large negative magnetoresistance in our earlier low-field (≤ 9T) measurements on TbPdBi [37], its origin remained mysterious; it was also unclear whether its magnetoresistance saturates in high field range. With the advantage of high-field measurements and theoretical analyses, we have an opportunity to address these issues in this work. Figures 2a-2b present the transverse and longitudinal magnetoresistivity MR (= $\frac{\rho_{xx}(B) - \rho_{xx}(0)}{\rho_{xx}(0)}$) of TbPdBi, measured with the current applied perpendicular and parallel to the magnetic field respectively at various temperatures (see the insets to Fig. 2a & 2b) (Note that these data were measured on the identical sample which was used for the Hall resistivity measurements shown in Fig. 1a). Both transverse and longitudinal MR decreases steeply with increasing magnetic field, and then tend to saturate above 15T; the magnitude of the MR's drop reaches ~98% as the field rises to 15T. Such a giant negative MR is observed at both $T \leq T_N$ and $T > T_N$. The magnitude of MR remains nearly temperature independent below 20K, but gradually decreases with increasing temperature above 20K (Fig. 2c). Even as the temperature rises to 200K, the negative MR remains significant, with its magnitude being ~ 20% at 9T (Fig. 2c). The MR changes from negative to positive only at room temperature, with the magnitude of positive MR being much smaller than that of negative MR at lower temperatures. Since TbPdBi is a superconductor with $T_c$ = 1.7K, its MR data measured at 1.7K (the base temperature of the measurement system) first shows a steep increase as the superconducting state is suppressed by magnetic field, then followed by the steep decrease as discussed above (see supplementary Fig. S5). We chose to not include this data in Fig. 2a-2b to avoid complications. It is worth pointing out that giant negative MR observed in TbPdBi does not occur to isostructural compounds (Gd/Tb/Dy)PtBi [10-12,38,39], which again suggests TbPdBi has a distinct electronic state from (Gd/Tb/Dy)PtBi.

Negative MR has been observed in various material systems and originates from several different mechanisms. In topological Weyl semimetals, the topological current induced by the chiral anomaly could lead to large negative MR when the magnetic field is parallel to the current, as observed in GdPtBi [11,38] and TbPtBi [10,12]. When the magnetic field is rotated away from the current, the chiral anomaly is gradually suppressed so that the sign of MR could change from negative to positive above a certain rotation angle. Prior studies have also shown materials with macroscopic disorders may also cause negative longitudinal MR, e.g. polycrystalline $Ag_{2+\delta}Se$ [40], gallium arsenide Quantum wells [41], and disordered topological insulator $TlBi_{0.15}Sb_{0.85}Te_2$ [42]. Another mechanism is the spin-scattering suppression driven by spin flip/flop transition; for instance, the colossal MR in $Ca_3Ru_2O_7$ can be attributed to this mechanism [43]. Apparently, none of the above mechanisms is relevant to our observation of giant negative MR in TbPdBi. This is because that (i) our observed negative MR is nearly independent of field orientation and tends to saturate in the high field regime, which excludes the chiral anomaly effect, (ii) our samples are high-quality single crystals and do not involve macroscopic disorders, and (iii) the giant negative MR of TbPdBi occurs at both $T \leq T_N$ and $T > T_N$. If its giant negative MR was due to the spin-scattering suppression in the spin flop transition, we would expect its MR's magnitude should significantly drop as the spin-flop transition is suppressed at $T > T_N$, which is inconsistent with the observation of the temperature independence of MR between $T_N$ and 20K and the survival of large negative MR in a wide temperature region above 20K (Fig. 2c). Additionally, we note that large negative transverse/longitudinal MR was recently reported in EuMnSb$_2$ [44] and EuB$_6$ [45]. The origin of large negative MR in EuMnSb$_2$ is ascribed to the field-induced metal-insulator transition, while the large negative MR of EuB$_6$ is suggested to be from the charge carriers' high spin polarization arising from a half Weyl semimetal [45]. As to be shown below, our band structure calculations for TbPdBi also reveals a high spin-polarization state, which is caused by a unique FM HTS state close to the topological critical point. Our observed giant negative MR in TbPdBi provides strong support for such a HTS state.

To understand the unusual transport properties of TbPdBi discussed above, we now present our detailed theoretical analysis of the electronic and transport properties of TbPdBi. Figures 3a-c show the arrangement of atoms in nonmagnetic, ferromagnetic (FM) with ordering vector [100], and A-type AFM state with propagation vector [111]. The high-symmetry nonmagnetic state of TbPdBi is described by the inversion asymmetric space group $T_d^2$ ($F\bar{4}3m$, No. 216). On considering the [100] FM state, the symmetry of the lattice reduces to fourfold rotary reflection $\bar{S}_4$, which combines the fourfold rotational and mirror symmetries. The associated band structure is semimetallic as shown in Fig. 3d without spin-orbit coupling. Importantly, the band structure constitutes a band inversion between $\Gamma_8$ (p-type) and $\Gamma_6$ (s-type) states reminiscent of half-Heulser materials only in the spin-down states whereas spin-up states remain uninverted. Such a unique band structure with single spin-channel band inversion can be closely associated with the half-metallic state of magnets and thus termed as HTS here. This HTS state can constitute multiband anticrossing points comprised of opposite spin channels. In the presence of SOC, the band anticrossing points are gapped, inducing a non-zero Berry curvature field and leading to large AHE [see Fig. 3e]. Figure 3f shows the band structure of the A-type AFM state of TbPtBi (ground state). The associated crystalline symmetries are threefold rotational symmetry ($C_3$) about [111] direction and spacetime symmetry $\tilde{T} = \left\{T \middle| \frac{1}{2}\frac{1}{2}\frac{1}{2}\right\}$, which combines both time-reversal symmetry and half-translation along [111] direction. The band structure is semimetallic with various band crossings along the $\Gamma - Z$ line and constitutes band inversion in both the spin-channels. This nontrivial band inversion AFM state evolves to an HTS state with a single-spin channel band inversion in presence of the magnetic field. Such a band structure evolution with the magnetic field can lead to an intermediate spin-canted state with various band anticrossing points, enhancing the Berry curvature field and associated AHE.

We want to emphasize that the distinct anomalous Hall resistivity peak of TbPdBi occurs near 5T (Fig.1a) where the system is in a spin-canted AFM state. We can therefore expect an additional contribution to the AHE stemming from the canted AFM state. As noted in the introduction, the non-collinear spin texture could induce large non-vanishing Berry curvature. Since our TbPdBi possess the same magnetic structure as that of GdPtBi/TbPtBi [46], i.e., the magnetic moments of Tb order ferromagnetically on the (111) planes, but are antiferromagnetically coupled along the [111] direction. When the magnetic field is applied long [100] direction, the magnetic moments will gradually tilt toward this direction, resulting in a spin-canted state. A recent theoretical study indicates that, in AFM nodal line materials $AMnBi_2$ (A=Ca, and Yb), a strong AHE could be induced by a weak spin canting, and the anomalous Hall conductivity keeps growing as the canting angle increases [47]. The remarkable anomalous Hall resistivity peak near 5T in TbPdBi probably is likely accredited to the canted spin state formed by Tb$^+$ ions. Such an interpretation is supported by our Berry curvature calculation at spin canted state, as discussed below.

To calculate the band structure and anomalous Hall conductivity (AHC) in the spin-canted states of TbPdBi under an external magnetic, we employ the virtual crystal approximation (VCA) with AFM and FM states model Hamiltonians as end members. The VCA Hamiltonian $H_{\text{vca}}$ of the spin canted state is defined as follows,

$$H_{\text{vca}} = (1-x)H_{\text{AFM}} + xH_{\text{FM}},$$

where $H_{\text{AFM}}$ ($H_{\text{FM}}$) is the Hamiltonian for AFM (FM) phase and $x$ is the tuning parameter from 0 to 1. $x = 0$ (1) denotes purely AFM (FM) phase while $0 < x < 1$ describes the spin-canted states. In the ground state ($x = 0$), the dominant Berry curvature $\Omega_{yz}$ resides on anticrossing points in the ΓZ direction as shown in Fig. 4a. With an increase of $x$ to 0.1 (low external magnetic field), the strong Berry curvature $\Omega_{yz}$ although remains on the ΓZ path, the finite FM coupling splits the oppositely spin-polarized bands, shifting the Berry curvature hot-spots to distinct energies [Fig. 4b]. In the fully polarized FM phase, the Berry curvature $\Omega_{yz}$ around Γ becomes significantly small [see Fig. 4c]. Calculated AHC for $x = 0 \sim 1$ as a function of energy is shown in Fig. 4d. The AHC is nearly zero at $x = 0$ due to Berry curvature cancellation induced by effective $\tilde{T}$ symmetry. A dramatical peak emerges near 0.1 eV above the Fermi level at $x = 0.1$ because the band splitting caused by breaking both C$_3$ and $\tilde{T}$ reduces the Berry curvature cancellation. The AHC peak gradually flattens and shifts to a higher energy with increasing $x$. The negative AHC at the Fermi level grows monotonically from $x = 0$ to 1, which implies that there are non-zero Berry curvatures away from Γ point in the spin-canted AFM states.

We consider the doping effect on the AHC behavior in the magnetic transition. For each doping concentration, the chemical potential is determined by fixing the number of occupied states. Positive (negative) doping concentration $\Delta n$ indicates the number of occupied states above (below) the original occupied particle number, i.e., $\Delta n = n-n_0$, where $n$ ($n_0$) is the current (origin) occupied particle number per formula unit cell. When $\Delta n = 0.0085$/f.u., $\sigma_{yz}$ reaches a maximum near $x = 0.1$, as shown in Fig. 4e, qualitatively consistent with our experimental observation (Fig.1a). Note that the VCA cannot fully capture the behavior of magnetic state evolution under an external magnetic field due to the lack of actual correlated-electronic corrections effects. In this sense, $x$ is not quantitatively comparable to the external field applied during the experiment. However, this approach can give a reasonable physical insight into the band structure effects in spin-canted states. On comparing to the experimental AHC (Fig. S4b), our calculated AHC are smaller than the experimental value. Such a discrepancy further suggests that effects other than those not captured in our VCA calculations are also at play.

Next, we first discuss the origin of the large Berry curvature in our system. Fig. 3d-e show the band structure without and with SOC when the magnetic field is applied along [100] direction. The band degeneracy lifting by exchange interaction is commonly seen in other Heusler compounds, such as GdPtBi/TbPtBi [9,11,12]. However, the weaker exchange interaction of Tb ions and nearly zero band inversion strength lead to the band splitting in TbPdBi smaller than that in (Gd/Tb)PtBi. As a result, the spin-split bands create more crossing points near the Fermi-level. A small gap is opened at the anti-crossing points with SOC, and the large non-vanished berry curvature is accumulated here.

We note that in REPtBi (RE = Gd, Nd), the non-vanishing Berry curvature is induced by the field-induced Weyl state. The location and number of Weyl nodes depend on the direction of applied magnetic field $B$. For TbPdBi, our calculation indeed reveals a Weyl state. We found that two pairs of Weyl nodes appear around $\Gamma$ point in the FM phase driven by the magnetic field applied along [100] axis; the positions of these Weyl nodes are summarized in supplementary table I. However, these Weyl nodes are far away from the Fermi level, which cannot induce large AHE. Furthermore, the absence of chiral anomaly revealed in our magnetotransport measurements indicates that the Weyl nodes do not contribute to the transport properties of TbPdBi. Hence, we believe the giant AHC in TbPdBi cannot be attributed to the Weyl state.

The second origin of the large Berry curvature is the spin-canted state. As revealed by our calculations, increasing the canted angle widens the band splitting and enlarges the gap at the anti-crossing points, which results in the Berry curvature along yz direction ($\Omega_{yz}$) continually increasing. However, the Berry curvature arises from the anti-crossing points and the spin-canted state is not the unique property for TbPdBi. It also has been observed in other half-Heusler compounds such as GdPtBi [9,11], TbPtBi [12], and DyPtBi [39]; their anomalous Hall conductivity and Hall angle are much smaller than that in TbPdBi. One may wonder what makes TbPdBi so unique. Our above band structure calculations reveal TbPdBi shows a distinct topological state: TbPdBi is at the critical point at the AFM ground state since its band inversion strength is nearly zero. At the field-driven FM state, it evolves to a HTS state: its spin-down bands maintain band inversion, while its spin-up bands lose their band inversion and become topological trivial. We note a similar state has been predicted in $EuB_6$ [48], where a large AHE and large negative MR have been recently reported [45]. Such findings imply that the HTS state probably plays a crucial role in enhancing the AHE in topological materials. As we stated before, the net Berry curvature requires the sum-up of the Berry curvature among all the occupied states. Thus, the number of the bands crossing the Fermi level could affect the Berry curvature cancellation. In general, the compounds showing good-metal behavior usually exhibit a large Berry curvature cancellation, resulting in a small net Berry curvature. However, in some compounds with bad-metal or semimetal behavior, only a few bands cross through the Fermi level, which could reduce the possibility of Berry curvature cancellation. In TbPdBi, the spin-up and spin-down bands are pushed to opposite directions, with a bandgap opening in the spin-up bands, which substantially reduces the number of the bands which cross the Fermi level, thus minimizing the Berry curvature cancellation.

The exotic HTS state in TbPdBi leads to not only a large AHE but also a giant isotropic negative MR. As described above, the spins are polarized near the Fermi level (spin-polarization reaches 40%) when applying an external magnetic field, such that the carriers from spin-down bands dominate the transport properties of TbPdBi. In this case, spin-scattering is significantly suppressed across the spin-flop transition, which leads to steep resistivity drop with the increase of magnetic field. When the magnetic field reaches above 15T, the spins are fully polarized (Fig. 1b), so the MR remains constant. Since the large negative MR is attributed to such an exotic 3D band structure, it is expected to be independent of the field orientation, which is exactly what we observed in experiments (Fig. 2a & 2b). The temperature independence of MR($B$) between $T_N$(=5.2K) and 20K (Fig. 2a&2b) can also well understood in terms of the field-induced FM HTS. Although there is not a long-range spin-flop transition above $T_N$, short-range AFM should exist in a temperature range above $T_N$ at zero field so that a crossover-like transition from PM to FM should occur under high magnetic fields, which is indeed manifested in the magnetization data measured at 20K as shown in Fig. 2b. This indicates a FM HTS can extend to temperatures above $T_N$ under high magnetic fields. However, when the temperature is increased above 50K, thermal excitation would bring out the electrons from the spin-up bands, thus increasing the spin-scattering and leading to the decrease of the magnitude of negative MR as shown in Fig. 2c.

In summary, we have observed an unusual AHE effect with a surprisingly large anomalous Hall angle ($\tan\Theta^H \approx 2$) and a giant negative MR in TbPdBi. Our theoretical analysis indicates that these exotic transport properties originate from the unique HTS state of TbPdBi, dominated by the contribution of spin-down bands. The greatly enhanced AHE results from a significantly reduced Berry curvature cancellation between the spin-down and spin-up bands. Spin canting of the AFM state under magnetic field increases the number of band anticrossing points and widens the gaps, which in turn increases the Berry curvature hot spots near the Fermi level and accounts for the distinct anomalous Hall resistivity peak at low fields. We also find that the HTS state enhances the spin-polarization for the spin-down band in the FM phase, which explains the giant negative MR. Our study unveils a new pathway for generating an extremely large AHE by tuning the electronic structure and the magnetic state in half-Heusler materials. It thus advances the understanding of the interplay between topological state and magnetism and provides clues for materials design for spintronics and applications.

**Methods**

The single crystals of TbPdBi were synthesized using the Bi-flux method [49]. The starting materials of Tb, Pd, and Bi powders were mixed with a molar ratio of Tb: Pd: Bi=1:1:20, loaded into $Al_2O_3$ crucibles, and sealed in quartz tubes under high vacuum. The mixtures of source materials were heated to 1050°C in a crucible furnace and held at this temperature for 48 hours for homogeneously melting, then followed by slow cooling down to 700°C at a rate of 3° C /h. Cubic-shape single crystals of TbPdBi were obtained by removing the excess Bi flux through centrifuging. The structures and compositions of the grown crystals are confirmed by X-ray diffraction measurements and Energy-dispersive X-ray spectroscopy (EDS). Magnetoresistivity and Hall resistivity measurements were performed using a standard six-probe method in a Physical Property Measurement System (PPMS, Quantum Design). The high-field transport measurements were carried out at the National High Magnetic Field Laboratory (NHMFL) in Tallahassee. The magnetoresistivity and Hall resistivity data presented in this paper are obtained through symmetrizing and antisymmetrizing the longitudinal and transverse Hall resistivity data measured at positive and negative magnetic fields respectively. All the samples used for transport measurements were polished to rectangular shapes with dimensions of ~0.8mm $\times$ 0.7mm $\times$ 0.2mm, and the polished surfaces were parallel to the (001) plane. The current was applied to the [100] or [001] directions for both longitudinal resistivity and Hall resistivity measurements. Magnetization measurements were measured using the National High Magnetic Field Pulsed Field Facility at Los Alamos National Laboratory and the SQUID magnetometer (Quantum Design).

Electronic structure calculations were performed within the density functional theory framework using the Vienna ab-initio simulation package (VASP) based on the projector-augmented wave method [50-52]. The generalized gradient approximation (GGA) was employed to include the exchange-correlation effects [53] and an on-site Coulomb interaction was added for Tb $f$-electrons within the GGA+U scheme with $U_{\text{eff}} = 10$ eV. A plane wave cut-off energy of 500 eV was used and a $11 \times 11 \times 11$ Γ centered $k$-point mesh to sample the first Brillouin zones of cubic and trigonal unit cells associated with various magnetic ordering. Spin-orbit coupling effects were included self-consistently. Transport properties were calculated using a first-principles material-specific, effective tight-binding model Hamiltonian generated using the VASP2WANNIER90 interface [54]. Tb $d$ and $f$ orbitals, Pd $s$, $p$, and $d$ orbitals, and Bi $p$ orbitals were included in constructing the wannier functions.

**Acknowledgment**

The experimental part of this work is based upon research conducted at The Pennsylvania State University Two-Dimensional Crystal Consortium–Materials Innovation Platform (2DCC-MIP), which is supported by NSF Cooperative Agreement No. DMR-2039351. Z.Q.M. also acknowledges the support from NSF under Grant No. DMR 2211327. The work at Northeastern University was supported by the Air Force Office of Scientific Research under award number FA9550-20-1-0322 and benefited from the computational resources of Northeastern University's Advanced Scientific Computation Center (ASCC) and the Discovery Cluster. The work at the National High Magnetic Field Laboratory is supported by the NSF Cooperative Agreement No.DMR-1644779 and No. DMR-1157490 and the State of Florida. The work at TIFR Mumbai is supported by the Department of Atomic Energy of the Government of India under Project No. 12-R&D-TFR-5.10-0100.

**Author contributions**

The crystal growth and transport measurements were carried out and analyzed by Y.L.Z, Y.W., & Z.Q.M. The high magnetic field measurements were carried out by Y.L.Z., D.G. S.H.L., L.J.M., J.C.P. & J. S. The theoretical work was done by C.Y.H., H.L., B.S. & A.B. This work is supervised by Z.Q.M. (experiment) and B.S. and A.B. (theory).

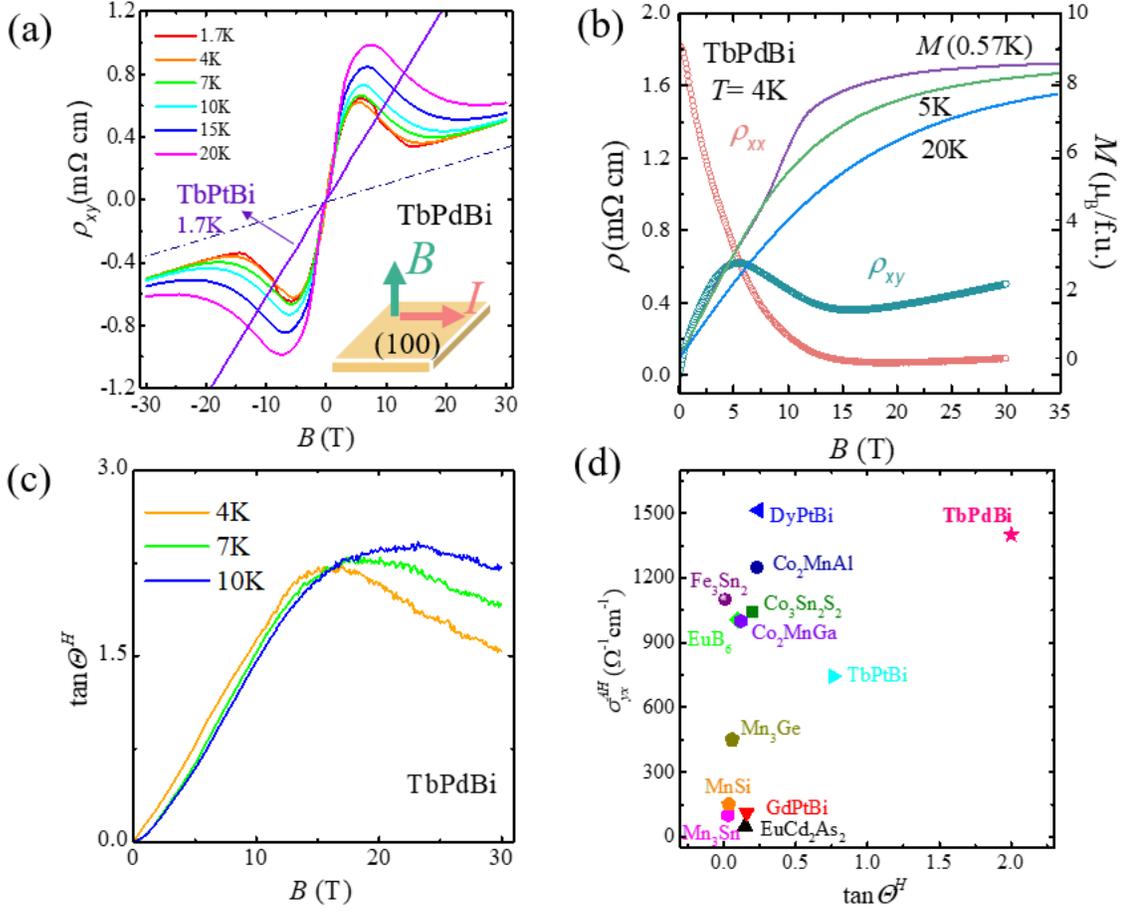

Figure 1. (a) Hall resistivity of TbPdBi as a function of the magnetic field $B$ up to 31T at various temperatures. The Hall data have been anti-symmetrized to remove the minor component of $\rho_{xx}$. Inset: the schematic of the experimental setup for Hall measurements. (b) Left axis: longitudinal resistivity $\rho_{xx}$ and Hall resistivity $\rho_{xy}$ at 4K for TbPdBi (marked as hollow circles). Right axis: magnetization measured at various temperatures (solid line). (c) The $\tan\Theta^H$ of TbPdBi as a function of magnetic field $B$ at 4K (below Neel temperature), 7K and 10K (above Neel temperature). (d) The comparison of $\sigma_{yx}^{AH}$ and $\tan\Theta^H$ between TbPdBi and other magnetic conductors [4,6,8,9,12,14,17,18,39,45,55-57].

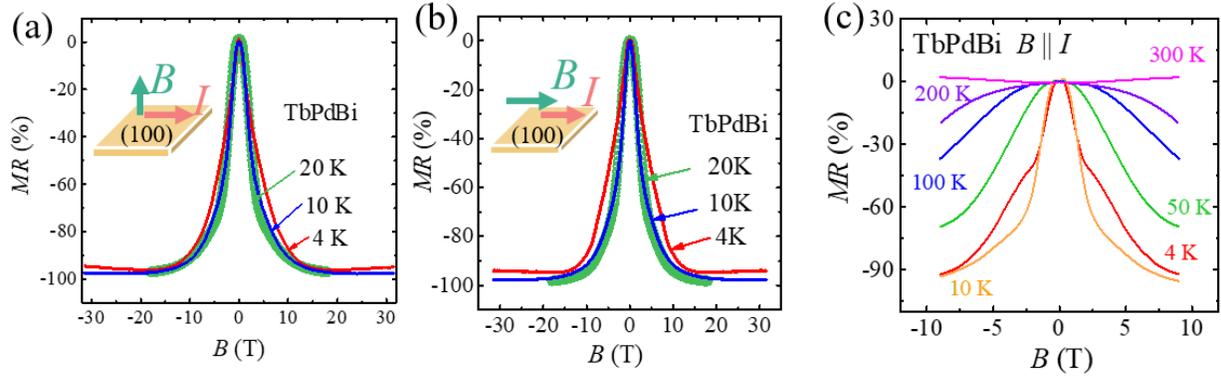

Figure 2. Magnetoresistivity (MR) of TbPdBi: (a-b) Transverse (a) and longitudinal (b) MR = $\Delta\rho/\rho_0 = [\rho(B)-\rho(B=0)]/\rho(B=0)$ at 4K (below Neel temperature), 10K and 20K (above Neel temperature). The MR at 4K and 10K were measured up to 31T, while the MR for 20K was measured up to 18T. The Insets show the schematic of the experimental setup for transverse and longitudinal MR measurements. (c) Longitudinal MR versus magnetic field measured in PPMS at various temperatures.

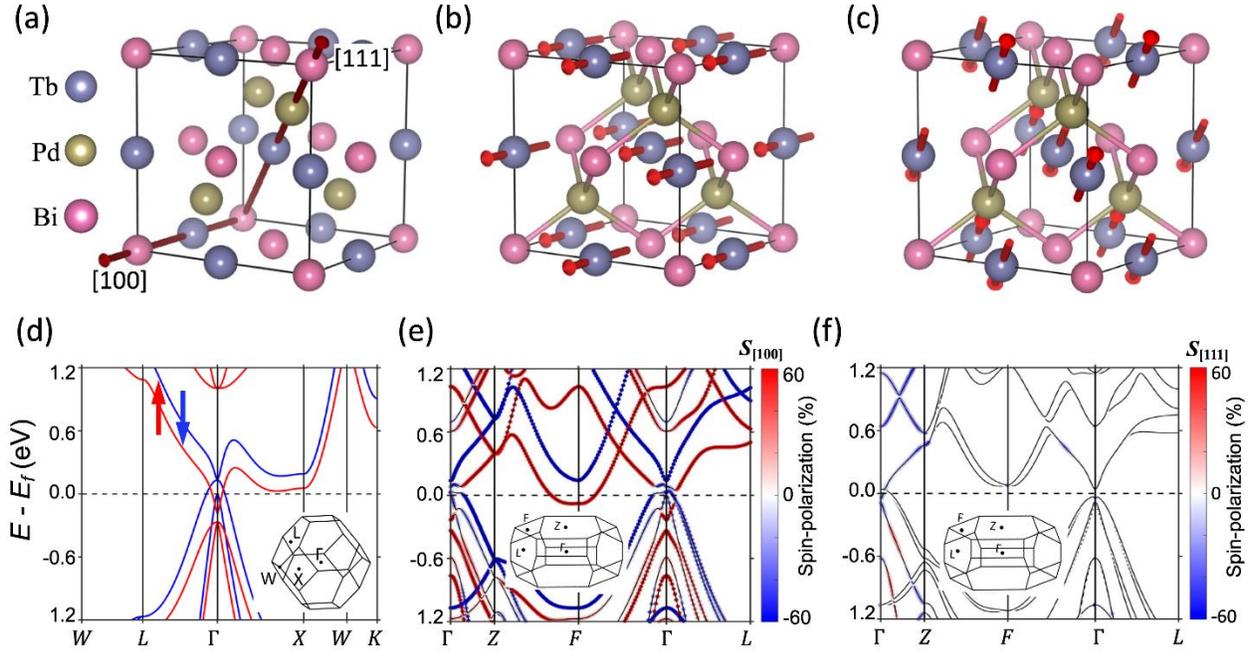

Figure 3. (a) Crystal structure of half-Heusler TbPdBi. [111] and [100] vectors denote the principal magnetic axes for AFM and FM states, respectively. Spin structures of (b) [100] FM and (c) [111] AFM magnetic structure in FM. Spins are aligned ferromagnetically in a (111) plane and antiferromagnetically between the (111) planes in the AFM state shown in (c). (d) Calculated spin-resolved bulk band structure of FM TbPdBi without SOC considering the primitive unit cell. (e) Spin-polarized band structure with SOC in (e) FM and (f) AFM states. Trigonal supercells are considered in (e) and (f). Insets in (d)-(e) show the first Brillouin zones associated with the unit cells employed in the calculations.

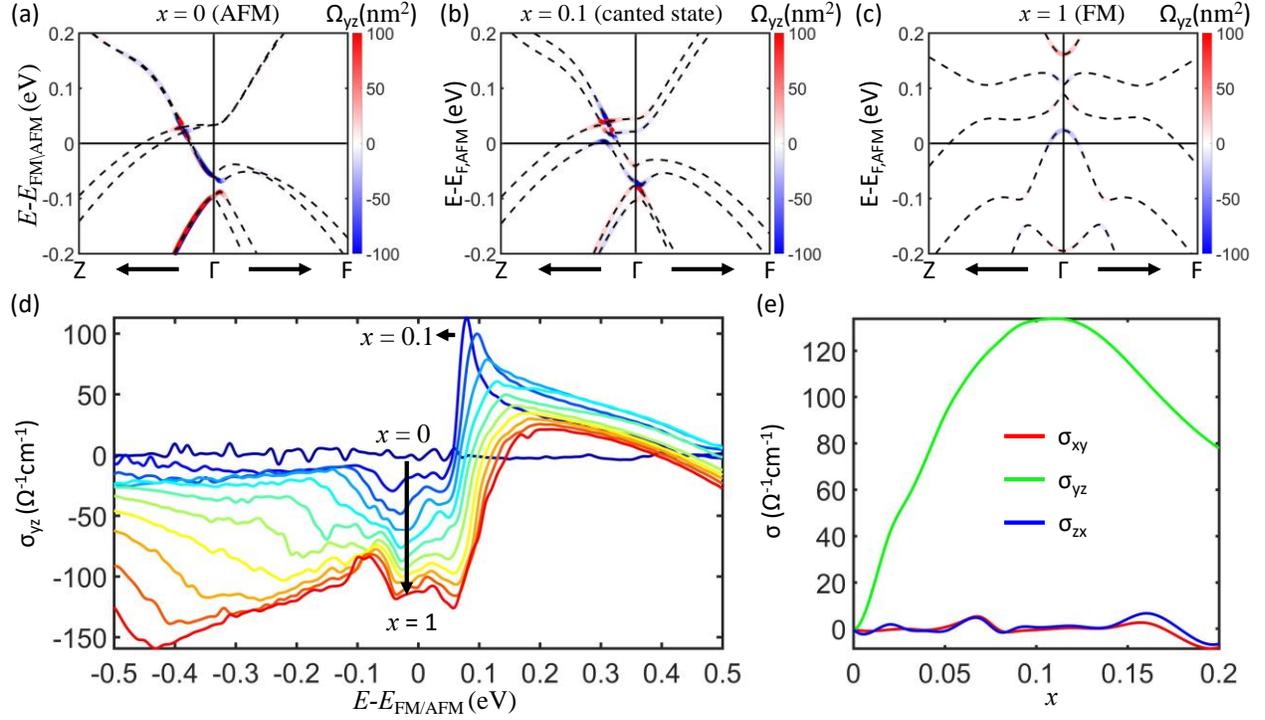

Figure 4. Berry curvature $\Omega_{yz}$ distribution of TbPdBi band structures for (a) $x = 0$ (AFM), (b) $x = 0.1$ (spin canted states) and (c) $x =1$ (FM) along selected high-symmetry directions. (d) Calculated anomalous Hall conductivity $\sigma_{yz}$ as function of energy for varying $x = 0$ (dark blue curve) to $x = 1$ (red curve). (e) Anomalous Hall conductivity as a function of $x$ with $\Delta n = 0.0085$/f.u. See the text for more details.